\documentclass[conference]{IEEEtran}
\usepackage{amsmath,amssymb,amsfonts}
\usepackage{algorithmic}
\usepackage{graphicx}
\usepackage{textcomp}
\usepackage{xcolor}
\usepackage{xspace}
\usepackage{subcaption}
\usepackage{booktabs}
\usepackage{tikz}

\usepackage[
  bibstyle=ieee,
  citestyle=numeric,
  isbn=true,
  doi=true,
  url=true,
  sorting=none,
  bibencoding=utf8,
  backend=biber,
  maxnames=99,        
]{biblatex}
\usepackage[linesnumbered]{algorithm2e}
\SetAlCapSkip{0.5em}

\begin{document}
\newcommand{\mypar}[1]{%
  \textbf{#1. }%
}
\newcommand{\mymod}[0]{%
  ~\mathrm{mod}~%
}
\newcommand{\digit}[3][\strut]{%
  {#2\vphantom{#1}}_{#3}%
}

\newcommand{\myspace}{\quad}

\def\R{{\mathbb{R}}}
\def\N{{\mathbb{N}}}
\newcommand{\spmvm}{SpMVM\xspace}

\title{Fast Entropy Decoding for Sparse MVM on GPUs}

\author{
\IEEEauthorblockN{Emil Schätzle}
\IEEEauthorblockA{\textit{Department of Computer Science} \\
\textit{ETH Zurich}\\
Zurich, Switzerland \\
emil.schaetzle@inf.ethz.ch}
\and
\IEEEauthorblockN{Tommaso Pegolotti}
\IEEEauthorblockA{\textit{Department of Computer Science} \\
\textit{ETH Zurich}\\
Zurich, Switzerland \\
tommaso.pegolotti@inf.ethz.ch}
\and
\IEEEauthorblockN{Markus Püschel}
\IEEEauthorblockA{\textit{Department of Computer Science} \\
\textit{ETH Zurich}\\
Zurich, Switzerland \\
pueschel@inf.ethz.ch}
}

\newcommand\copyrighttext{%
  \footnotesize \textcopyright 2026 IEEE.  Personal use of this material is permitted.  Permission from IEEE must be obtained for all other uses, in any current or future media, including reprinting/republishing this material for advertising or promotional purposes, creating new collective works, for resale or redistribution to servers or lists, or reuse of any copyrighted component of this work in other works.\newline
  Published in: 2026 IEEE International Parallel and Distributed Processing Symposium (IPDPS), pp. 416--432, DOI: 10.1109/IPDPS65963.2026.00044.}
\newcommand\copyrightnotice{%
\begin{tikzpicture}[remember picture,overlay]
\node[anchor=south,yshift=10pt] at (current page.south) 
  {\fbox{\parbox{\dimexpr\textwidth-\fboxsep-\fboxrule\relax}{\copyrighttext}}};
\end{tikzpicture}%
}


\maketitle
\copyrightnotice

\begin{abstract}
We present a novel, practical approach to speed up sparse matrix-vector multiplication (\spmvm) on GPUs. The novel key idea is to apply lossless entropy coding to further compress the sparse matrix when stored in one of the commonly supported formats. Our method is based on dtANS, our new lossless compression method that improves the entropy coding technique of asymmetric numeral systems (ANS) specifically for fast parallel GPU decoding when used in tandem with \spmvm. We apply dtANS on the widely used CSR format and present extensive benchmarks on the SuiteSparse collection of matrices against the state-of-the-art cuSPARSE library. On matrices with at least \boldmath$2^{15}$ entries and at least 10 entries per row on average, our compression reduces the matrix size over the smallest cuSPARSE format (CSR, COO and SELL) in almost all cases and up to 11.77 times. Further, we achieve an \spmvm speedup for the majority of matrices with at least \boldmath$2^{25}$ nonzero entries. The best speedup is $3.48\times$. We also show that we can improve over the AI-based multi-format AlphaSparse in an experiment that is limited due to its extreme computation overhead. We provide our code as an open source C++/CUDA header library, which includes both compression and multiplication kernels. 
\end{abstract}


\section{Introduction}
\label{sec:introduction}

Sparse matrix–vector multiplication (\spmvm) is a fundamental linear algebra primitive that is the key building block in a wide range of applications in scientific computing, machine learning, and beyond. Examples include iterative solvers of linear equations~\cite{baker2005}, finding eigenvectors of a matrix~\cite{golub1977,grimes1994}, and more recently, inference in (pruned) neural networks, including large language models (LLMs)~\cite{frantar2023sparse, dettmers2023spqr}. In these domains, \spmvm frequently constitutes the bottleneck of entire pipelines, dominating the overall runtime.
This makes the availability of \spmvm-kernels with the highest performance a critical research goal.

\spmvm is inherently memory-bound, and thus the key to performance is efficient representation of the sparse matrix, storing only nonzero elements, and reducing the amount of data movement overall. The most popular example is the compressed sparse row (CSR) format, which offers excellent spatial locality. Starting with the early work on blocking CSR~\cite{saab1994sparskit} many other formats have been proposed to further compress the matrix beyond CSR if it has additional structural properties~\cite{spmv2002,patternspmv2009,Willcock2006,Kourtis2010} (see also the related work in Section~\ref{sec:related_work}) but most have not found support in current libraries.

GPUs have become a natural platform for \spmvm kernels, due to their parallel design and high memory bandwidth. But also on GPUs, \spmvm is memory-bound, and increasingly so, due to the overarching hardware trends~\cite{memwall2024}. The main design principle for the matrix format and its \spmvm implementation is to optimize memory transfer by storing as little data as possible while ensuring full bandwidth utilization through coalesced memory reads and balanced work among streaming multiprocessors (SMs). The current CUDA library~\cite{cusparse} supports seven formats including CSR, COO, and SELL, and each has strengths and weaknesses. AlphaSparse~\cite{Du2022} proposes an AI-based solution to predict, for a given matrix, the best format, and thus \spmvm runtime, among those and several others, but the overhead for doing so is prohibitive for most applications.

One avenue for further performance gains is lossy matrix compression, which is used in LLMs by quantizing the matrix entries to low-bit precision while minimizing the penalty on the model's quality (called perplexity)~\cite{frantar2023gptq, lin2024awq}. However, in many other domains, including scientific computing, preserving the full content of information, and thus precision, is critical.

\mypar{Contributions} 
In this paper, we are, to the best of our knowledge, the first to use lossless entropy coding to compress sparse matrices beyond what is possible with current formats. Doing so yields a memory-optimal representation of the distribution of indices and values that define the matrix. We then provide a GPU kernel that performs \spmvm while decoding the compressed matrix on the fly.

The main technical contribution is the entropy coding method, called dtANS, which we develop. It is a novel adaptation of the existing asymmetric numeral system (ANS)~\cite{Duda2013}, specifically designed for fast decoding on GPUs so as not to be a bottleneck in \spmvm.

Our dtANS can be applied to any existing sparse matrix format and, in fact, input data in other memory-bound GPU applications. We will focus on CSR as a starting point.
\begin{itemize}
    \item We present dtANS, a novel form of entropy coding, adapted from the existing ANS~\cite{Duda2013}, that enables fast decoding on GPUs.
    \item We use dtANS to provide a novel solution for high performance \spmvm on GPUs that operates with an entropy-coded sparse matrix format obtained by applying dtANS to CSR. We call this format \emph{CSR-dtANS}. Thus, our \spmvm is performed interleaved with the on-the-fly decoding of the matrix.
    \item More precisely, before applying dtANS to CSR, we also perform delta-encoding of the indices in CSR to further reduce their entropy in many cases. Using random graph models, we provide some evidence for why this works.
    \item We extensively evaluate CSR-dtANS on the SuiteSparse Matrix Collection~\cite{Davis2011}. On matrices with at least $2^{15}$ nonzeros and at least 10 nonzeros per row on average, our compression reduces the matrix size over the best cuSPARSE format (smallest of CSR, COO, and SELL) in almost all cases and by up to 11.77 times. Further, we achieve a speedup for the majority of matrices with at least $2^{25}$ nonzero entries. The best speedup is by $3.48 \times$. We also show that we can improve on the AI-based multi-format AlphaSparse~\cite{Du2022} in an experiment that is limited due to the extreme computation overhead of AlphaSparse. 
    \item We provide our code as an open source C++/CUDA header library, which includes both compression and multiplication kernels. 
\end{itemize}

\section{overview}

\begin{figure}
\centering
\includegraphics[width=\columnwidth]{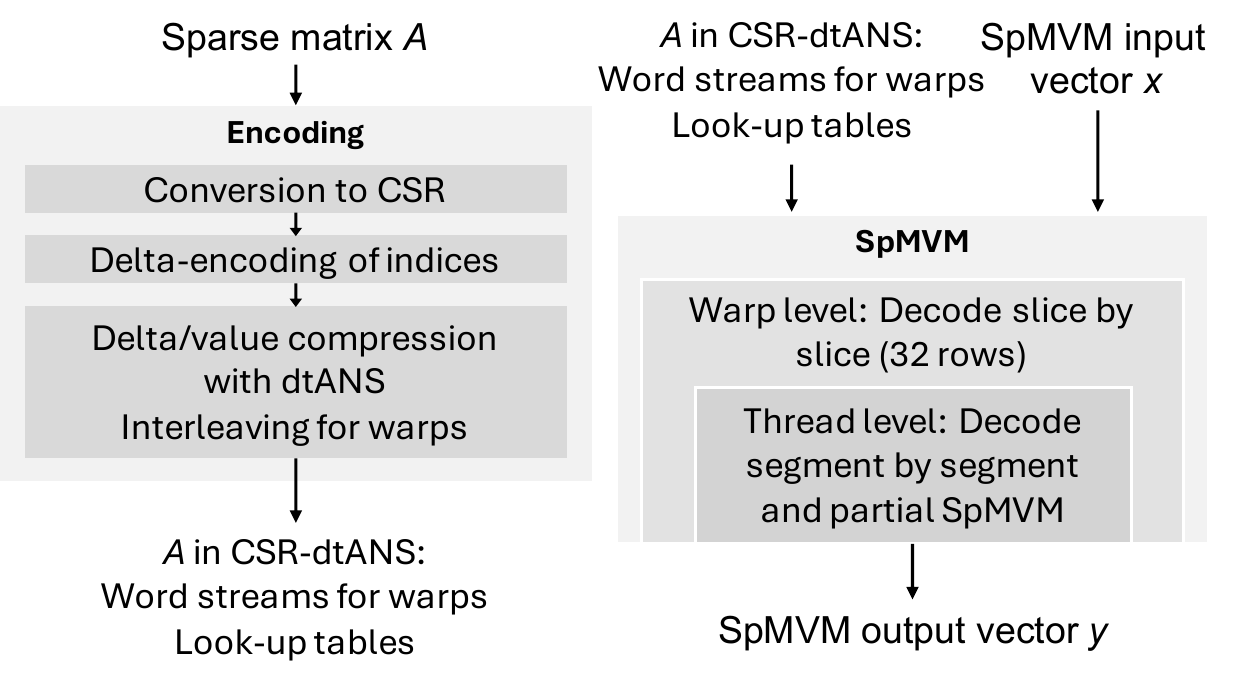}
\caption{Encoding into CSR-dtANS (left) and \spmvm on CSR-dtANS (right)}
\label{fig:overview_block_diagram}
\end{figure}

We provide a high level overview of our \spmvm toolchain. First, we outline the steps necessary for encoding, which is a one-time effort for each sparse matrix \(A\). Second, we sketch our \spmvm kernel that decodes \(A\) for each dense vector \(x\). A block diagram of both processes can be found in Fig.~\ref{fig:overview_block_diagram}. Details can be found in Section~\ref{sec:methods}. 

\subsection{Encoding}
We describe the four steps of compressing the sparse matrix \(A\) to our proposed representation.

\mypar{Input} As input we take the sparse matrix \(A\). In our implementation, the matrices are read from \texttt{.mtx} files.

\mypar{Conversion to CSR} In a first step, the matrix is converted to the CSR format, which we chose for its popularity, simplicity, and generality. More advanced formats introduce complexity to improve performance for specific classes of matrices. This exceeds our scope of demonstrating the applicability of on the fly entropy decoding for memory-bound compute kernels. However, our subsequent entropy coding (and specifically dtANS) could be used with other formats.

\mypar{Delta-encoding of indices} Next, we apply delta-encoding~\cite{Kourtis2010} to the indices. This stores the differences between ascending column indices of the same row rather than the indices themselves, which typically reduces the entropy of the resulting distribution. For a more detailed discussion, we refer to Subsection~\ref{par_methods_delta_encoding}.

\mypar{Delta/value compression with dtANS} Afterward, the deltas and values obtained are compressed row-by-row using our entropy coding method \emph{dtANS}. This creates look-up tables shared by all threads and a data stream of 4-byte words for each thread. Each of these encodes a row of \(A\).

\mypar{Interleaving for warps} Memory accesses must be \emph{coalesced} for efficiency, which means that all reads in a group of 32 threads or \emph{warp} target consecutive addresses. Therefore, we interleave the streams of threads in a warp.

\mypar{Output} The CSR-dtANS compressed representation of the matrix consists of the look-up tables and the warp-interleaved word streams. The encoded data can be stored in memory or saved in a file for repeated decoding.

\subsection{Decoding and \spmvm} 
 We outline the decoding process.
 
\mypar{Input}
As input we take the CSR-dtANS representation of \(A\) consisting of the look-up tables and warp-interleaved word streams as well as a dense vector \(x\) as input to \spmvm.

\mypar{SpMVM}
Each warp processes a \emph{slice} of 32 consecutive rows with consecutive threads operating on consecutive rows. The \spmvm kernel operates accordingly slice by slice on the matrix. Each thread processes the elements in segments, meaning that 4 nonzeros of a segment are processed in parallel. The obtained nonzeros are used on the fly to compute their product with the vector \(x\). After all elements have been processed, the results are written to the output.

 \mypar{Output}
 The result of the \spmvm kernel is \(y = Ax\).

\section{Background}
\label{sec:background}

We provide the necessary background on sparse matrix-vector multiplication (\spmvm) and the most common storage formats used in current libraries: coordinate list (\emph{COO}), compressed sparse row (\emph{CSR}), and sliced ellpack (\emph{SELL}) format. Then we explain compression through entropy coding with an emphasis on \emph{tANS}, which is the basis of our compression method for fast \spmvm on GPUs.

\subsection{Sparse Matrix-Vector Multiplication}

A matrix $A\in\R^{m\times n}$ is sparse if it has few nonzero elements. Given vectors $x,y\in\R^m$, \spmvm computes
$$
y = Ax + y,
$$
typically done in floating-point arithmetic.

\mypar{Sparse formats}
The key to performance in \spmvm is efficient storage of the matrix $A$.
The \emph{coordinate} format (COO)~\cite{Hwu2023} stores the nonzeros in three arrays: row indices, column indices, and values. Accordingly, two indices are stored for each nonzero. The \emph{compressed sparse row} format (CSR)~\cite{Im2004} stores values and column indices in row major order; thus a third array only needs to store the start index of each row in those. An example can be found in Fig.~\ref{fig:background_csr}. CSR is the starting point for our work.

\begin{figure}
\centering
\[
\begin{bmatrix}
  & 7 &   & 5 \\
3 &   & 2 &   \\
  & 4 &   &   \\
  &   &   & 1 \\
\end{bmatrix}
\Leftrightarrow
\begin{array}{ll}
    \texttt{Values}         & 7, 5, 3, 2, 4, 1 \\
    \texttt{Column Indices} & 1, 3, 0, 2, 1, 3 \\
    \texttt{Row Start}      & 0, 2, 4, 5, 6 \\
\end{array}
\]
\caption{Example for the CSR format with 6 nonzeros}
\label{fig:background_csr}
\end{figure}

CSR provides near perfect spatial locality for \spmvm, but comes with challenges for parallel processing on GPUs: first, it does not balance work well between different cores and second, it only provides memory coalescing when rows are processed by entire warps (requiring many nonzeros per row)~\cite{Hwu2023}. The sliced ELLPACK (SELL) format~\cite{Monakov2010} is specifically designed for SIMD including GPUs. It compresses groups of rows together as a slice. Rows in a slice are padded to match the number of nonzeros in the row with the most nonzeros and the result is stored in column-major order. This reduces the number of indices to one per slice and one per nonzero but introduces an additional padding overhead.

\mypar{Comparison}
The best format depends on the sparsity pattern of the matrix. In terms of compression, COO is best for matrices with very few nonzeros per row (because empty rows require no memory), SELL is best for matrices with a similar number of nonzeros per row (because no padding and fewer offsets are required), and CSR is best for matrices with a varying number of nonzeros per row (because it asymptotically requires a single index and no padding). The performance of \spmvm depends both on the compression and the memory access patterns. 

Our work applies a particular form of entropy coding to $A$ in CSR format for further compression and potential speedup of \spmvm.

\subsection{Entropy Coding}

Entropy coding refers to a family of lossless data compression techniques that aim to reduce the size of a corpus of data to its theoretical limit~\cite{Pearlman2011}. In particular, given a sequence of symbols (here these would be the values and indices in CSR) drawn independently from some probability distribution \(P\) (over \(\Omega\)),~\cite{Shannon1948} proves that the expected number of bits per symbol needed to represent such a sequence is lower bounded by its entropy $H$, defined by
\begin{align}
\label{eq:background_entropy}
H(P) &= \sum\limits_{\omega\in\Omega}-P(\omega)\cdot\log_2(P(\omega)).\\
\intertext{In entropy coding, \(P\) is approximated by \(P'\) and the average number of bits per symbol approaches the cross entropy \(H'\), defined by}
\label{eq:background_cross_entropy}
H(P, P') &= \sum\limits_{\omega\in\Omega}-P(\omega)\cdot\log_2(P'(\omega)).
\end{align}
Note that \(H(P) = H(P, P)\) and that \(H(P, P') \geq H(P)\) is smaller the closer \(P'\) approximates \(P\).
Two popular types of entropy coding are \emph{Arithmetic Coding}~\cite{Rissanen1976} and \emph{Huffman Coding}~\cite{huffman1952}. The former, real-based Arithmetic Coding accurately approximates the probability distribution of the symbols, resulting in better compression ratios and worse decompression time. The latter bit-based Huffman Coding roughly approximates the probability distribution of the symbols with powers of two, resulting in worse compression ratios and better decompression time. Both are not suitable for efficient on the fly decoding on GPUs.

The main theoretical contribution of this paper is a form of entropy coding that is suitable for GPUs and, when applied to the CSR format, enables fast on the fly decoding to speed up \spmvm. We derive it as a variant of \emph{Asymmetrical Numeral Systems} (ANS)~\cite{Duda2013}. ANS is integer-based and obtains compression ratios close to Arithmetic Coding and decompression time close to Huffman Coding. There are multiple variants, most important for us, \emph{tabled ANS} (tANS). ANS used mixed-radix representations of numbers that we explain first. Then we explain encoding and decoding in tANS in detail. Our work will then adapt tANS for use with \spmvm on GPUs in Section~\ref{sec:methods}.

\subsection{Mixed radix numeral systems}

Natural numbers are commonly represented using \emph{positional numeral systems} such as the decimal, binary, or hexadecimal system. In these systems, the base (or radix) is the same for all digits, e.g., $\digit{161}{10} = \digit{10100001}{2} = \digit{\textrm{A1}}{16}$, and the value of each digit is determined by its position in the number. However, a positional numeral system may not always be the most natural choice. For example, durations are commonly expressed as a combination of days, hours, and minutes. Each component functions similarly to a digit, albeit with different bases: there are 24 hours in a day, but 60 minutes in an hour. Thus, in a \emph{mixed radix numeral system}, each digit is annotated with a base. For example, 11370 minutes represented as 10 days, 5 hours, and 30 minutes would be expressed as: 
$$
\digit{11370}{10} = 10 \cdot 24 \cdot 60 + 5 \cdot 60 + 30 = \digit{10}{\infty}\digit{5}{24}\digit{30}{60}.
$$
The represented number is obtained by multiplying each digit by the product of the bases of the less significant positions and summing up the results. Thus, to represent arbitrarily large numbers, we also need to allow arbitrarily large numbers in the most significant position (i.e., base $\infty$). 

\begin{figure}
\centering
\includegraphics[width=0.8\columnwidth]{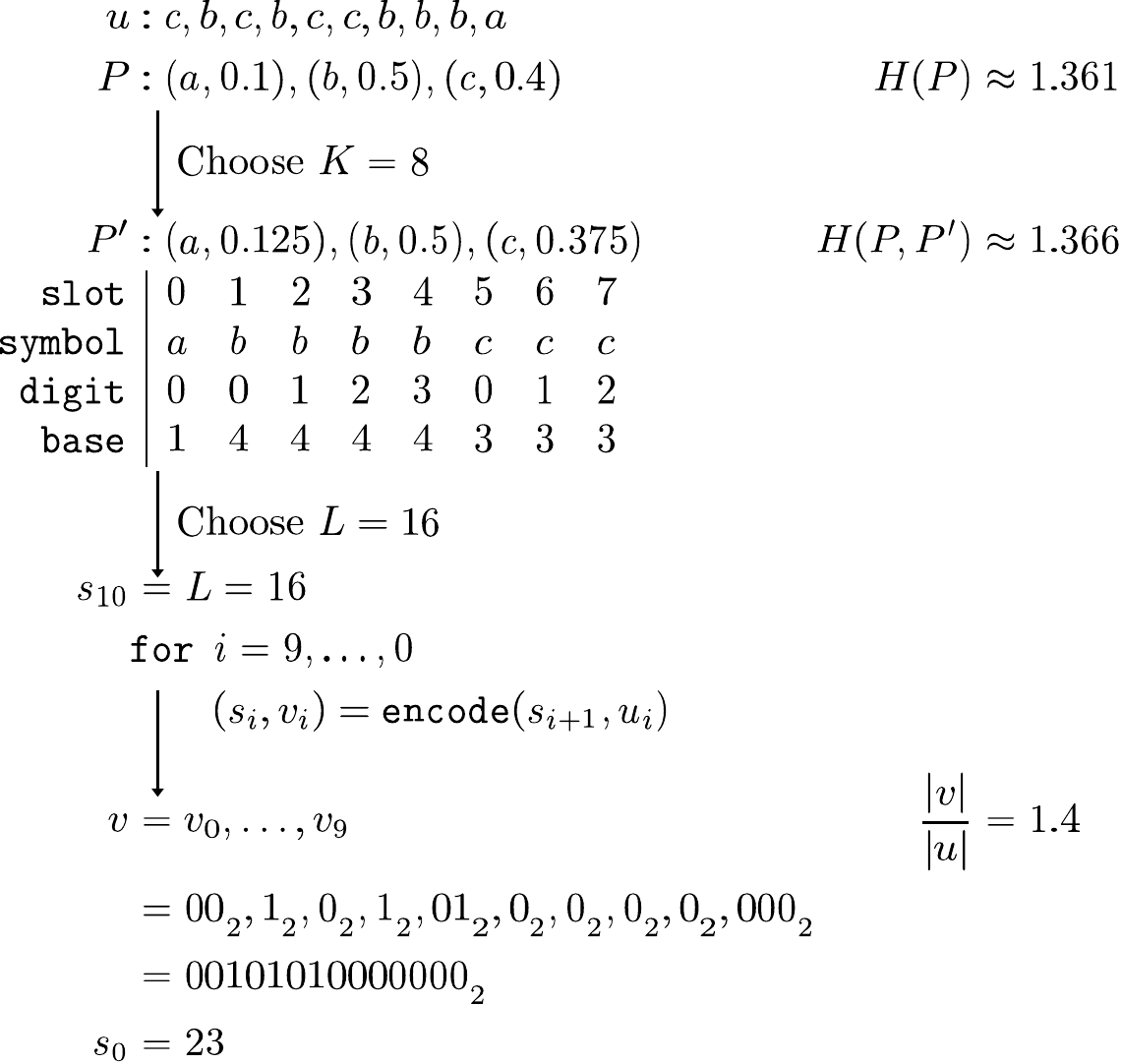}
\caption{Example of entropy encoding with tANS}
\label{fig:background_tANS}
\end{figure}

\subsection{Encoding with tANS}
\label{subsec:background_encoding_tans} 

We explain encoding with tANS~\cite{Duda2013}. We do so first with a representative small example and then provide pseudocode for the general case. The example and the high-level encoding steps are shown in Fig.~\ref{fig:background_tANS} and explained next. The input is a sequence $u$ of ten symbols; the encoded output $v$ is a bit stream that is stored together with the tables that enable decoding. In our \spmvm application, the input symbols would be integers (delta-encoded indices) or floating-point numbers (values). Note that in realistic applications, the needed tables would be negligible in size compared to $u$ and $v$.

\mypar{Determine symbol distribution}
We first compute the distribution $P$ of the $n=10$ input symbols in $u = (c,b,c,b,c,c,b,b,b,a)$, which we write here using pairs (symbols, probability). $P$: $(a, 0.1), (b, 0.5), (c, 0.4)$. There are $m=3$ symbols. The entropy, computed using \eqref{eq:background_entropy}, is $H\approx 1.361$ bits. This means that optimal entropy encoding would require $10\cdot H\approx 14$ bits for $v$.

\mypar{Size of coding tables and approximating \boldmath$P$}
Next, we decide on the size $K$ of the coding tables, which is recommended in $[2m,4m]$. Here we choose $K=8$. Then, each symbol is assigned a fraction of slots as shown in the \texttt{symbol} table in Fig.~\ref{fig:background_tANS}, resulting in another distribution $P'$: $(a, 0.125), (b, 0.5), (c,0.375)$. In effect, this moves the denominator $10$ in $P$ to the denominator $8$. $P'$ should be chosen to be as similar as possible to $P$; formally, this means minimizing the cross entropy $H'= H(P, P')$, computed using \eqref{eq:background_cross_entropy}, which here is $\approx 1.366$ bits and determines the achievable compression with tANS, which thus is close to the one given by $H$. For example, choosing $P''$: $(a,0.25), (b, 0.5), (c,0.25)$ would yield the suboptimal $H(P,P'') = 1.5$ bits.

\mypar{Constructing the coding table}
Now, we fill in the coding tables shown in Fig.~\ref{fig:background_tANS}. We put the symbols into the \texttt{symbol} table in any order and in consecutive slots according to the chosen distribution $P'$. In the \texttt{digit} table, equal symbols are numbered consecutively, and the \texttt{base} table contains the multiplicities of each symbol in $P'$. The \texttt{slot} table just stores the indices of the entries as shown.

\mypar{Preparing for encoding}
Now we can start to encode each symbol $u_i$ in $u$ into a bit string $v_i$. The encoding is done from right to left, so here we start with $u_9 = a$. The encoding maintains and uses an internal state $s$ in each step, i.e., it has the form $(s_i,v_i) = \text{encode}(s_{i+1},u_i)$, here $i = 9,\dots,1$. The bit strings are written to $v_i$ and $s$ stays \emph{normalized} in $\mathcal{L} = \left[L, 2L-1\right]$. Intuitively, $s$ stores information with sub-bit granularity. The parameter $L$ should be $\geq K$ and chosen as large as possible to reduce information loss while still allowing operations within a single instruction. We choose $L = 16$, i.e., $\mathcal{L} = [16,31]$. Next, we go through the first two encoding steps.

\mypar{Encoding \boldmath$u_{n-1}$}
The initial state is set as $s_{10} = L = 16$. We start by encoding $u_9 = a$ by computing $(s_9,v_9)=\texttt{encode}(16,a)$. Both are computed jointly, using different mixed radix numeral systems. We state the result before explaining the individual steps:
$$
    s_{10}=16=\digit{2}{\infty}\underbrace{\digit{000}{2}}_{b=v_9}\digit{0}{1} \quad\text{and}\quad\digit{2}{\infty}\digit{0}{8} = 16 = s_9.
$$
First, the radix $r$ of the lowest digit in $s_{10}$ is determined from the \texttt{base} table. For $u_9=a$ it is $r=1$. The associated digit is obtained as $(s_{10}\mymod r) = (16\mymod 1) = 0$. The latter $0$ is used for table lookup: as \texttt{digit}, which together with the \texttt{base} and \texttt{symbol} yields the \texttt{slot} number of $0$ in this case. This 0 with radix $K = 8$ (table size) yields the least significant digit $\digit{0}{8}$ for $s_9$. Finally, we write $s_{10}$ in the form $\digit{x}{\infty} \digit{b}{2} \digit{0}{1}$, where $b$ is a long enough bit string so that $\digit{x}{\infty} \digit{0}{8}$ (which will be $s_9$) is $\in\mathcal{L}$. Here $b = \digit{000}{2}$, which is $v_i$, and $x=2$, which yields $s_9$ as shown.

\mypar{Encoding  \boldmath$u_{n-2}$}
We continue encoding \(u_8 = b\) by computing $(s_8,v_8)=\texttt{encode}(16,b)$: 
$$
    s_{9}=16=\digit{2}{\infty}\underbrace{\digit{0}{2}}_{b=v_8}\digit{0}{4} \quad\text{and}\quad \digit{2}{\infty}\digit{1}{8} = 17 = s_8.
$$
For $b$, \texttt{base} yields $r = 4$, $(16\mymod 4 = 0)$, so the least significant digit for $s_9$ is $0_4$. Table lookup with $\texttt{digit} = 0$, $\texttt{base} = 4$ and $\texttt{symbol} = b$ yields \texttt{slot} number 1. So $s_8 = \digit{x}{\infty} \digit{1}{8}$. This time only one bit $\digit{0}{2}$ is needed such that $s_9 = \digit{x}{\infty} \digit{0}{2} \digit{0}{4} = \digit{2}{\infty} \digit{0}{2} \digit{0}{4}$ and $s_8 =  \digit{2}{\infty} \digit{1}{8} = 17\in\mathcal{L}$. Thus $v_8 = \digit{0}{2}$.

Note how the encoding operates as expected: infrequent symbols like $a$ get more bits (3 in this case) and frequent ones like $b$ fewer (only 1 in this case). Intuitively, this is because we have more slots to choose from for frequent symbols, allowing us to embed more information into this choice.

\mypar{Encoding: generic case}
Algorithm~\ref{alg:background_tans_encode} formalizes the generic decoding procedure.

\begin{algorithm}[t]
\SetKwInOut{Input}{Input}
\SetKwInOut{Output}{Output}
\Input{$u$, (\texttt{symbol}, \texttt{digit}, \texttt{base}), $n$}
\Output{$s_0$, $v$}
$s = L$\\
\For{$i=n-1$ \KwTo $0$}{
    $(s,v_i) = \texttt{encode}(s,u_i)$
}
\Return $s$, $(v_0,\ldots,v_{n-1})$

\SetKwFunction{FEncode}{\texttt{encode}}
\SetKwProg{Fn}{Function}{:}{}
\Fn{\FEncode{$s$, $u_i$}} {
    $r = \texttt{base}$ of $u_i$\\
    $d = s\mymod r$\\
    Choose slot $j$ for $\texttt{symbol}=u_i$, $\texttt{digit}=d$ and $\texttt{base}=r$\\
    Rewrite $s$ as $\digit{x}{\infty} \digit{b}{2} \digit{d}{r}$ such that $\digit{x}{\infty}\digit{j}{K} \in \mathcal{L}$\\
    $s = \digit{x}{\infty}\digit{j}{K}$\\
    \Return $s$, $b$
}
\caption{Encoding $u$ in tANS}
\label{alg:background_tans_encode}
\end{algorithm}

Applied to $u$ it yields the final state $s_0 = 23$ and the following output bit stream
$v$ in the order from right to left: 
$\digit{00}{2},
\digit{1}{2},
\digit{0}{2},
\digit{1}{2},
\digit{01}{2},
\digit{0}{2},
\digit{0}{2},
\digit{0}{2},
\digit{0}{2},
\digit{000}{2} = 
\digit{00101010000000}{2}$. 
The latter has 14 bits and is thus optimal.

The encoded representation of $u$ includes $v$, the final state $s_0=23$, and the three coding tables.

\subsection{Decoding with tANS}

Next, we explain the decoding with tANS, which is done by reversing the steps in the encoding. We start with the first decoding step of the example in Fig.~\ref{fig:background_tANS} and then provide pseudocode for the general case.

\mypar{Decoding \(u_0\)}
The decoding is performed in reverse order starting from \(s_0=23\). We rewrite 
\[
    s_0 = 23 = \digit{2}{\infty}\digit{7}{8} \quad\text{and}\quad \digit{2}{\infty} \underbrace{\digit{00}{2}}_{v_1}\digit{2}{3}=26=s_1.
\]
The radix of the lowest digit in $s_{0}$ is fixed as \(K=8\). The digit can be obtained as $(s_{0}\mymod K) = (23\mymod8) = 7$ and is interpreted as the address of a slot. The slot gives us a digit/base pair \(\digit{2}{3}\) and the symbol \(u_0=c\). Finally, we write $s_{1}$ in the form $\digit{2}{\infty} v_1\digit{2}{3}$, where the length of $v_1$ in the bit stream $v$ is recovered from requiring $s_1 \in\mathcal{L}$.

\mypar{Decoding: generic case}
Similarly, we can decode \(u_1,\ldots,u_9\) iteratively. Algorithm~\ref{alg:background_tans_decode} formalizes the process. We obtain \(u\) in its original order.

\begin{algorithm}[t]
\SetKwInOut{Input}{Input}
\SetKwInOut{Output}{Output}
\Input{$s_0$, $v$, (\texttt{symbol}, \texttt{digit}, \texttt{base}), $n$}
\Output{$u$}
$s = s_0$\\
\For{$i=0$ \KwTo $n-1$}{
    $(s,u_i)=\texttt{decode}(s,v)$ \label{line:background_tans_decode_step}
}
\Return $(u_0,\ldots,u_{n-1})$

\SetKwFunction{FDecode}{\texttt{decode}}
\SetKwProg{Fn}{Function}{:}{}
\Fn{\FDecode{$s$, $v_i$}} {
    $j = s \mymod K$\\
    $d = \texttt{digit[}j\texttt{]}$\\
    $r = \texttt{base[}j\texttt{]}$\\
    $s = \digit{x}{\infty} \digit{b}{2} \digit{d}{r}$ with $b \gets v$ such that $s\in\mathcal{L}$

    \Return $s$, $\texttt{symbol[}j\texttt{]}$
}
\caption{Decoding $u$ in tANS}
\label{alg:background_tans_decode}
\end{algorithm}

\section{\spmvm with dtANS}
\label{sec:methods}

We explain how we use entropy coding to further compress a CSR-represented matrix $A$ to achieve (depending on $A$) speedups for \spmvm on GPUs. This is done by decoding $A$ on the fly while executing \spmvm as shown in Fig.~\ref{fig:overview_block_diagram}. The key to \spmvm performance is fast decoding.

Our presentation follows Fig.~\ref{fig:overview_block_diagram}. First, we present delta-encoding, which further reduces the achievable entropy of CSR, and is done before entropy coding. Then we give a high level view of our proposed CSR-dtANS format and explain the shortcomings of tANS for on the fly-decoding. Finally, we present the details of the dtANS decoding and encoding and explain the implementation.

\subsection{Delta-encoding}
\label{par_methods_delta_encoding}

Many sparse matrices exhibit structure, such as consecutive elements, blocks, or clusters.
We exploit the fact that \emph{delta-encoding}~\cite{Kourtis2010} indices can reduce entropy for many of those. The idea is simple: instead of storing the indices in CSR, their differences are stored. We delta-encode rows separately.

\begin{figure}
\centering
\includegraphics[width=\columnwidth,trim={0 0.3cm 0 1.3cm},clip]{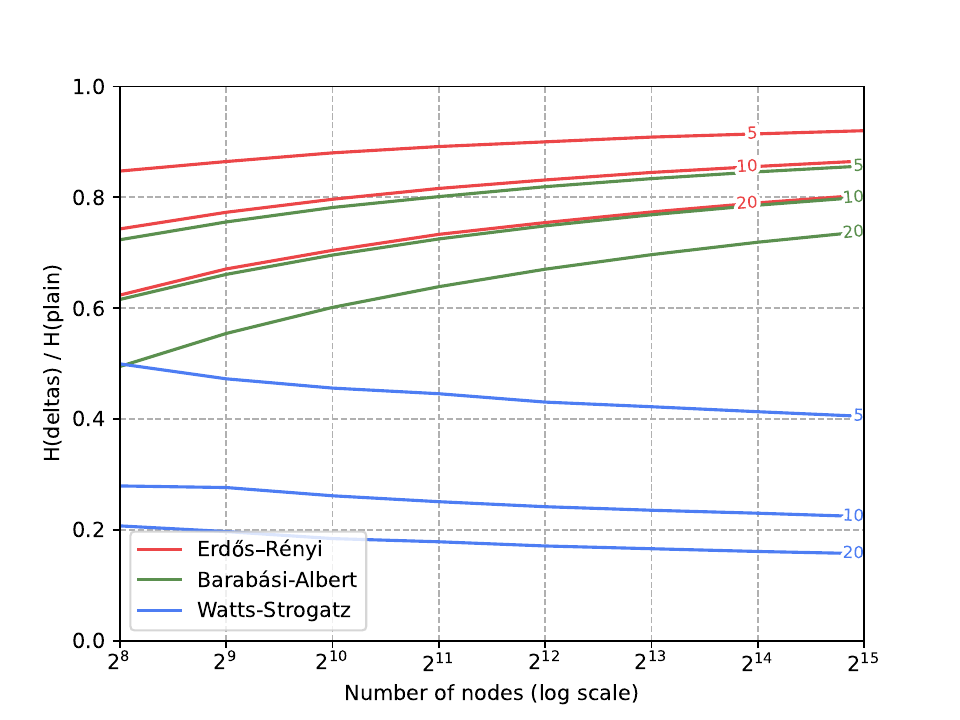}
\caption{Entropy reduction via delta-encoding for three random graph models with increasing number of nodes. Model parameters are chosen to keep the average degree at 5, 10, and 20.}
\label{fig:graph_models_entropy}
\end{figure}

For example, in tridiagonal matrices, the delta column indices would contain two 1s and one value between 0 and $n-1$. Similarly, only few values would be obtained for matrices with a $2$- or $3$-dimensional nearest neighbor relation such as stencils. It also works for common random graph models. For example, the Erdős–Rényi random graph model~\cite{Erdos1959} chooses every edge with a fixed probability $p$, which creates a geometrically descending distribution of deltas. Fig.~\ref{fig:graph_models_entropy} shows the entropy reduction on the indices for three common random graph models: Erdős–Rényi, Watts-Strogatz~\cite{Watts1998}, and Barabási-Albert~\cite{Barabasi1999}, which produces so-called small-world graphs. Three different parameter choices are considered for each model to keep the average degree at 5, 10, and 20, respectively. The x-axis increases the number of nodes and the y-axis shows the relative entropy achieved, which is reduced in all cases. We report the median of three runs to reduce variance.

\subsection{CSR-dtANS}

\begin{figure}
\centering
\includegraphics[width=\columnwidth]{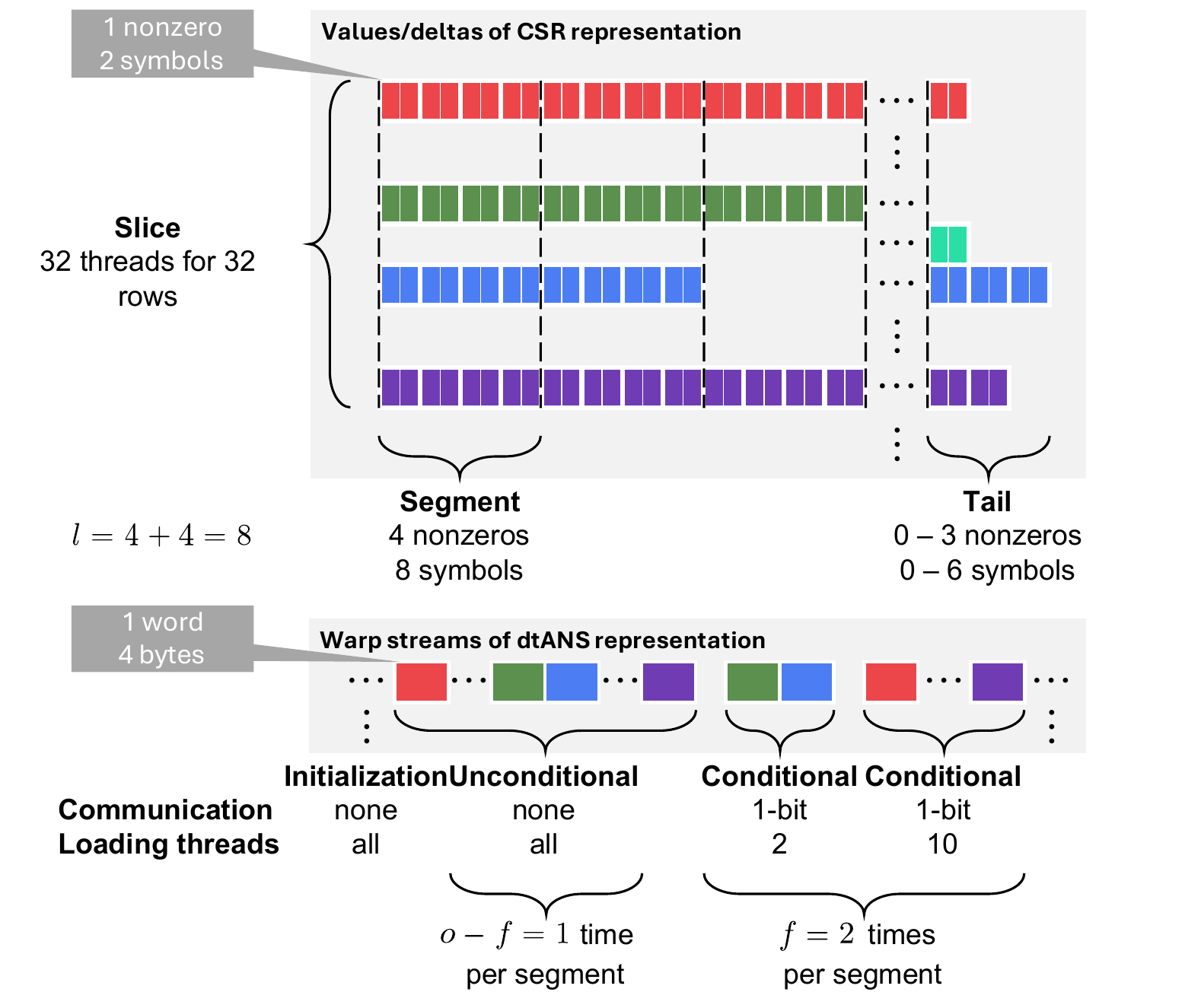}
\caption{Mapping of data to threads and dtANS representation}
\label{fig:methods_dtANS}
\end{figure}

We start with a high-level explanation of the intended memory layout and mapping of work to GPU execution units of our proposed entropy-encoded sparse matrix format \emph{CSR-dtANS} following Fig.~\ref{fig:methods_dtANS}. Entropy coding is applied after delta-encoding the indices in CSR as explained above. Specifically, we identify the main shortcomings that tANS exhibits for fast GPU decoding and how we resolve them with dtANS.

\mypar{Work distribution}
We map a \emph{slice} of 32 consecutive rows to a \emph{warp} of 32 threads one-by-one. To supply each thread with the required \emph{symbols} for computing a single dot product of the \spmvm, we sort the nonzeros of the CSR representation by increasing column index, delta-encode the column indices, and break the values/deltas into rows while interleaving them, such that value and delta of each nonzero are adjacent. This stream is then compressed using entropy coding.

\mypar{Lack of efficient SIMT parallelism for warp processing}
Entropy coding provides a binary stream for each thread. However, reading these streams in parallel is not efficient on a GPU because the memory accesses are not to consecutive addresses or not \emph{coalesced} causing low memory throughput. Thus, we want to appropriately interleave the binary streams obtained for each thread of a warp. This requires communication at the warp level for every decoded symbol, as threads have to compute the prefix sum of the bits loaded by each thread to obtain their individual offsets from a common stream position. In dtANS, the compressed binary stream $v$ stores 4-byte words instead of individual bits. This requires less frequent synchronization as more data is loaded at once. Additionally, the required prefix sum is only over integers 0 and 1, which can be efficiently implemented using the intrinsic \texttt{\_\_ballot \_sync} once and \texttt{\_\_popc} twice.

\mypar{Lack of ILP for efficient thread execution}
The decoding steps of tANS have a sequential dependency on the state $s$ (Line~\ref{line:background_tans_decode_step} of Algorithm~\ref{alg:background_tans_decode}), which destroys instruction level parallelism (ILP). dtANS resolves this by grouping $l=4+4$ consecutive symbols required for 4 nonzeros into \emph{segments}. The decoding then proceeds in parallel for symbols within each segment but sequentially across segments. This increases ILP, leads to better utilization of execution units, and coins the name \emph{decoupled} tANS. As the length of the rows might not be a multiple of 4, special treatment is required for the \emph{tail} (see Section~\ref{subsec:methods_implementation_details}). In CSR-dtANS, for each segment, \(o=3\) words of 4 bytes are required, of which \(f = 2\) are loaded \emph{conditionally} with intra-warp communication and \(o-f=1\) are loaded \emph{unconditionally} without intra-warp communication. We explain how we derive these parameter choices in Subsection~\ref{subsec:methods_decoding_dtANS}.

\subsection{Table construction in dtANS}
\label{subsec:methods_decoding_dtANS}

As in tANS, the compressed representation of \(u\) consists of coding tables, an encoded stream \(v\), and the length $n$ of \(u\). The coding table, which is shared by all threads, is obtained as in tANS by approximating the symbol distribution of the entire matrix to \(K\) slots. Thus, its size is K times the size of an entry and does not depend on the number of encoded symbols. However, we introduce a new parameter \(M\) that upper-bounds the multiplicity of each symbol in the approximations. In tANS \(M = K\) which is also a valid choice in dtANS. On the one hand, a small \(M\) increases the achievable cross-entropy since some previously viable approximations are excluded. Intuitively, the information stored in a slot along with the encoded symbol is limited, making frequent symbols more expensive to encode. On the other hand, a small \(M\) increases the synchronicity of the data consumption of the threads in a warp making more loads \emph{unconditional} as all threads are guaranteed to consume data at a certain minimum rate.

\subsection{Decoding in dtANS}

For dtANS, obtaining the output stream \(v\) is more involved than for tANS because it is optimized for fast parallelized decoding. Thus, we will first present decoding following the example in Fig.~\ref{fig:background_tANS} and then provide pseudocode for the general case. As some parameters used for dtANS are not practical for demonstration purposes, we present dtANS for a machine with a word-size of 2 bits. We choose \(K = 8\) as before and introduce \(M = 4\) which \(P'\) already fulfills allowing us to reuse the coding tables.

\mypar{Words in \boldmath$v$}
The compressed representation \(v\) of \(u\) is \(\digit{11123121100}{4}\). The radix of this stream is \(4\), which means that each \emph{word} in \(v\) is between $0$ and $3$ (i.e., 2 bits). We define an \emph{unpacking function} \texttt{unpack} that takes 3 words \(w_1\), \(w_2\) and \(w_3\) and returns 2 slots \(i_1\) and \(i_2\) between $0$ and $7$ by rewriting 
\[
\digit{i_2}{8}\digit{i_1}{8} = \digit{w_1}{4} \digit{w_2}{4} \digit{w_3}{4}.
\]
As this is a bijection, the inverse \texttt{pack} exists.
In the general case, let \(W\) be the radix of \(v\) and \(o\) be the number of words \texttt{unpack} takes to output \(l\) slots. Large \(W\) reduce losses due to imprecision and minimize the synchronization overhead. However, we also want all operations to be performed in a single instruction. Thus, a natural choice is the size of a register on the given machine. As \texttt{unpack} must be surjective to make every combination of slots representable, we choose \(o\) minimal such that \(K^l\geq W^o\). If this is not an equality, some information will be lost during unpacking. For our example, equality holds as \(8^2 = 64 = 4^3\). The precise implementation of \texttt{pack} and \texttt{unpack} is a design choice. But for powers of two, the rewriting demonstrated above is a good choice because it breaks down to changing the interpretation of a bitstring. In CSR-dtANS we choose \(W=2^{32}\) as the word size of the GPU. We choose \(K=2^{12}=4096\) such that the coding tables are large enough for many symbols but still fit in shared memory. Now we can choose \(o=3\) as in our small example and obtain equality of \(K^l\geq W^o\) for \(l=8\) which provides sufficient instruction level parallelism.

\mypar{Preparing for decoding}
We start by reading 3 words from \(v\): \(w_1 = 1\), \(w_2 = 1\) and \(w_3 = 1\). In the general case, these are \(o\) words \(w_1,\ldots,w_o\). We also initialize the internal state of the decoder by setting the digit \(d=0\) and the radix \(r=1\).

\mypar{Decoding \boldmath$(u_0, u_1)$}
Now, we use \texttt{unpack} to obtain slots \(i_1=5\) and \(i_2=2\) as \[\digit{1}{4} \digit{1}{4} \digit{1}{4} = 21 = \digit{2}{8}\digit{5}{8}.\]
They are used for lookups into \texttt{symbol}, \texttt{digit}, and \texttt{base}. The former will give us the encoded symbols \(u_0 = c\) (for \(i_1\)) and \(u_1 = b\) (for \(i_2\)). The latter two will give us two returned digit/base pairs \(\digit{0}{3}\) (for \(i_1\)) and \(\digit{1}{4}\) (for \(i_2\)). In the general case, we obtain \(l\) slots \(i_1,\ldots,i_l\) from \(w_1,\ldots,w_o\) which result in \(l\) symbols and \(l\) returned digit/base pairs. Observe how the individual symbols can be decoded and processed in parallel. In the next paragraph, we will explain how these digit/base pairs are then accumulated to the decoder state. 

\mypar{Obtaining \boldmath$w_1,w_2,w_3$ for the next segment}
We eventually have to reconstruct \(w_1\), \(w_2\), and \(w_3\) for the decoding of the next segment. Therefore, we add the returned digits to the right of the decoder state \(\digit{0}{1}\).
We start by adding the first returned digit \(\digit{0}{3}\) getting \[\digit{0}{1}\digit{0}{3}=\digit{0}{3}.\] Next, we check whether the information stored in the decoder state suffices to extract a word which is not the case as \(r=3<4=W\) resulting in reading \(w_1=\digit{2}{4}\) from \(v\). This is the first \emph{conditional} load because it depends on the internal state of the decoder. We continue by adding the second returned digit \(\digit{1}{4}\) giving \[\digit{0}{3}\digit{1}{4}=\digit{1}{12}.\] Now, \(r=12\geq 4=W\) such that we can extract \(w_2=d\mymod W=1\mymod4=1\) and update \[d = \bigg\lfloor\frac{d}{W}\bigg\rfloor=\bigg\lfloor\frac{1}{4}\bigg\rfloor=0\quad\text{and}\quad r= \bigg\lfloor\frac{r}{W}\bigg\rfloor=\bigg\lfloor\frac{12}{4}\bigg\rfloor=3.\]
This is the second conditional load. The flooring of \(r\) in the update causes some information to be lost, and thus larger \(W\) reduce these losses. \(w_3 = 3\) is read from \(v\) \emph{unconditionally} by all threads completing the three words \(w_1=2\), \(w_2=1\) and \(w_3=3\) for the next iteration.
In the general case, we need to provide \(w_1,\ldots,w_o\) for the next step. Therefore, we check the decoder state during \(f \leq l\) conditional loads potentially extracting a word if \(r \geq W\) and loading from \(v\) otherwise. Afterwards, we load the remaining \(o-f\) words from \(v\) unconditionally. The placement of these checks is a design choice we can optimize for latency, size of intermediate numbers, and data dependencies. To guarantee that the radix of our decoder state stays below a constant upper-bound, we need \(M^l \leq W^f\) such that all returned digits can be used instead of loading from \(v\) during conditional loads. At the same time, the number of checks \(f\) can be at most the total number of required words $o$ giving \(l\cdot\log_W(M) \leq f \leq o\). Ideally, the left inequality should be equal again to use instructions optimally. For our example, \(2\leq f\leq3\) is fulfilled by \(f=2\) and equality holds. In CSR-dtANS, we choose \(M=2^8=256\) which results in returned digits of at most 8 bits, a convenient size for efficient computation (Section~\ref{subsec:methods_implementation_details}). Now we can choose \(f=2\) as in our small example and also obtain equality.

\mypar{Decoding: generic case}
Similarly, we can decode \((u_2,u_3),\ldots,(u_8,u_9)\) iteratively. Algorithm~\ref{alg:methods_dtans_decode} formalizes the process. We obtain \(u\) in its original order.

\begin{algorithm}[t]
\SetKwInOut{Input}{Input}
\SetKwInOut{Output}{Output}
\Input{$v$, (\texttt{symbol}, \texttt{digit}, \texttt{base}), $n$}
\Output{$u$}
$w_1,\ldots,w_o \gets v$\\
$d = 0$\\
$r = 1$\\
\For{$j=0$ \KwTo $\frac{n}{l} - 1$ }{
    $i_{l \cdot j},\ldots,i_{l \cdot (j + 1) - 1} = \texttt{unpack}(w_1,\ldots,w_o)$\\
    \tcc{This loop can be parallelized and interleaved with the next}
    \For{$k = l \cdot j$ \KwTo $ l \cdot (j + 1) - 1$}{
        $u_k = \texttt{symbol[}i_k\texttt{]}$\\
        $\digit{d}{r} = \digit{d}{r}\digit{\texttt{digit[}i_k\texttt{]}}{\texttt{base[}i_k\texttt{]}}$
    }
    \For{$k = 1$ \KwTo $f$}{
        \uIf{$r \geq W$}{
            $w_k = d \mymod W$\\
            $d = \big\lfloor\frac{d}{W}\big\rfloor\quad\text{and}\quad r = \big\lfloor\frac{r}{W}\big\rfloor$
        }\Else{
            $w_k \gets v$\\
        }
    }
    \For{$k = f+1$ \KwTo $o$}{
        $w_k \gets v$
    }
}
\caption{Decoding $u$ in dtANS}
\label{alg:methods_dtans_decode}
\end{algorithm}

\subsection{Encoding in dtANS}

We only discuss encoding with dtANS on a high level due to lack of space. The details depend on the exact implementation of the decoder and do not provide additional insights. Again, we note that all code will be made available as open source. Encoding requires two passes over the data: in the forward \emph{base} pass it is determined after which checks the decoder needs to read information from \(v\) and in the backward \emph{digit} pass the exact words stored in \(v\) are determined. 

\mypar{Base pass}
We can determine the radix of the decoder state, and thus the branches taken during decoding without knowing the exact slots used to represent the symbols, as the \texttt{base} is identical for all slots associated with the same symbol. We store the branching information in a boolean array indicating which loads can be replaced by extracting a word from the decoder state. 

\mypar{Digit pass}
The previous base pass provided us with the control flow of the decoder in the form of a boolean array. The digit pass will go through the decoding steps backward to select the exact slots to represent the symbols and produce the words to store in \(v\) in lockstep. For going backwards, the order and direction of the loops as well as the placement of the checks have to be reversed exactly.

\subsection{Implementation Details of CSR-dtANS}
\label{subsec:methods_implementation_details}

We present various implementation details relevant for correctness and performance.

\mypar{Storing decremented radixes}
All radices in CSR-dtANS can be between 1 and some two-power depending on the data being encoded. To save one bit of storage if the actual upper limit has to be represented, we decrement the radices by one. In the computation, this can be compensated for by using an FMA to add the decremented value an additional time instead of a multiplication incurring no additional cost.

\mypar{Positioning of checks}
The previous tweak allows us to accumulate 4 returned digits (with a radix of up to \(M=2^8\)) into a 4-byte digit/base pair. This accumulation is a reduction and can benefit from some ILP while only requiring integer operations on words. Afterwards, we can add the returned digits to the decoder state using a word-size multiplication for the lower half of the result and the \texttt{\_\_umul\_hi} intrinsic for the upper half. 

\mypar{Escaping rare values}
There is no guarantee, that there are no more than \(K=4096\) different symbols of each domain calling for an escape mechanism to make them representable. Further, even if there are not too many symbols, the cross-entropy can be reduced significantly in some cases by escaping rare values (because assigning them a slot in the table is more expensive than paying the cost to escape them). To this end, a special symbol is introduced. Whenever it is decoded, an uncompressed symbol is read from the stream. Unfortunately, this requires synchronization after every decoding step for coalesced accesses hindering performance. An alternative would be to load these escaped values from a separate stream \textemdash{} trading more latency due to a bad memory access pattern for fewer instructions required. In CSR-dtANS, we approximate the exact distributions such that the expected total size is minimized.

\mypar{Efficient handling of end of row}
In the algorithm previously presented, the decoder has \(o=3\) words in the end that are known to be 0. However, since it knows the number of symbols \(n\), loading words that are known not to be needed can be skipped, reducing overhead. Moreover, should the length of the row not be divisible by 4, we can pad with any symbol which the decoder can then ignore as it knows \(n\).

\mypar{Tables in shared memory}
To maximize throughput and minimize latency of the lookups, we place the coding tables in shared memory. It is also not required to assign symbols consecutive slots in the tables. By randomly permuting them, we can reduce the number of accesses to the same \emph{memory-bank} (which are serialized) when decoding adversarial data.

\section{Results}
\label{sec:results}

We evaluate CSR-dtANS in terms of both the compression ratio achieved on sparse matrices and the performance of the associated \spmvm kernels. In particular, we compare CSR-dtANS with the formats COO, CSR, and SELL, implemented by the state-of-the art library cuSPARSE on all matrices in the SuiteSparse Matrix Collection~\cite{Davis2011}. We perform this comparison for warm and cold cache and 64- and 32-bit precision. Additionally, we conduct a limited experiment that compares CSR-dtANS to formats found by the machine-learning-based AlphaSparse autotuner~\cite{Du2022}.

We also considered the portable performance library Kokkos Kernels \cite{Rajamanickam2021}. Kokkos Kernels can call vendor implementations (cuSPARSE in our case), but also provide some native kernels. However, we found that the fastest cuSPARSE format outperforms their native kernels for medium-sized to large matrices (i.e., at least $2^{20}$ nonzeros) and rare speedups were negligible. For small matrices, Kokkos kernels could offer better gains, but, as we will see and is obvious from our approach, CSR-dtANS is not a good fit for those and already slower than the fastest cuSPARSE format. Thus, we decided against including Kokkos Kernels in the benchmarks.

\mypar{Experimental setup}
Experiments are conducted on an RTX 5090 GPU with 32 GB GDDR6 VRAM, a 96 MB L2 cache, and 170 streaming multiprocessors (SMs) with 128 KB L1 cache each. The host CPU features an Intel Xeon w5-2465X with 128 GB DDR5 ECC RAM, and a 33.75 MB L3 cache.
We use Ubuntu 24.04. with kernel 6.14.0 and the NVIDIA driver 575.57.08 (open) with cuSPARSE version 12.9. The environment provides nvcc 12.9.86 and gcc 13.3.0. We compile our code and cuSPARSE using nvcc-flags -O3 -Xptxas -O3 -std=c++20 -gencode arch=compute\_120,code=sm\_120 and gcc-flags -O3 -std=c++23 -march=native. We benchmark the performance using Nsight Compute 2025.2.1.0 (with clock control disabled). For CSR-dtANS and the cuSPARSE formats, we perform 7 runs and take the median runtime. For AlphaSparse, we use the runtime it reports.

\mypar{Evaluation dataset} From the 9030 sparse matrices in SuiteSparse, we consider those whose field type can be represented by a floating-point number (\emph{real}, \emph{integer}, and \emph{pattern}). This excludes the \emph{complex}-valued matrices and leaves 8975 matrices. Then, we convert all data types to floating point numbers and use 32-bit integer indices. Next, we compress the matrices using the formats CSR, COO, SELL, and our CSR-dtANS. Note, that with 32-bit indices cuSPARSE can only handle matrices with less than $2^{31}$ nonzeros. However, this does not limit cuSPARSE in our setting, as all its considered formats would exceed the available GPU memory for a matrix beyond that size.

\mypar{Cache state}
We provide both warm-cache and cold-cache measurements, since both are relevant depending on the use case. For example, iterative system solvers will likely run in a warm-cache setting as the code needs to read the same matrix multiple times. On the other hand, in the case of machine learning models, a new matrix is read for each layer.

\mypar{64- and 32-bit}
We provide results for 64- and 32-bit floating-point numbers. While 64-bit precision is the gold standard in scientific computing applications~\cite{Bailey2005}, other applications, such as machine learning models, can benefit from reduced memory footprints when using 32-bit precision. Note that the choice of precision only affects the size of the coding tables (and possibly escaped values) in CSR-dtANS and not the amount of bits to encode the symbols. Consequently, the 64-bit setting is generally more favorable for dtANS, as better compression can be expected.

\mypar{AlphaSparse} AlphaSparse uses a machine learning-based autotuning algorithm to choose the best sparse format representation to use. We use its most recent version\footnote{\url{https://github.com/PAA-NCIC/AlphaSparse}, commit \texttt{e12833f}}, available at the time of evaluation. It relies on specialized configuration files to guide its search. In these files, we raised the maximally achievable performance to enable the full potential of our GPU. We use the benchmarking infrastructure included in AlphaSparse's codebase to run its performance measurements\footnote{In some cases, the reported final performance is incorrect, so we contacted AlphaSparse's authors to extract the correct performance from the log.}. These benchmarks are executed with a warm cache. One major difference between our kernels and AlphaSparse's is that in the case of symmetric matrices, AlphaSparse does not duplicate off-diagonal elements, and instead only executes the multiplication with the lower or upper triangular half of the matrix. We modified our code accordingly to make the results comparable for those matrices.
\label{par:alphasparse}

\subsection{Compression achieved}

\begin{figure}
\centering
\begin{subfigure}{\linewidth}
\includegraphics[width=\columnwidth,trim={0 0 0 1.2cm},clip]{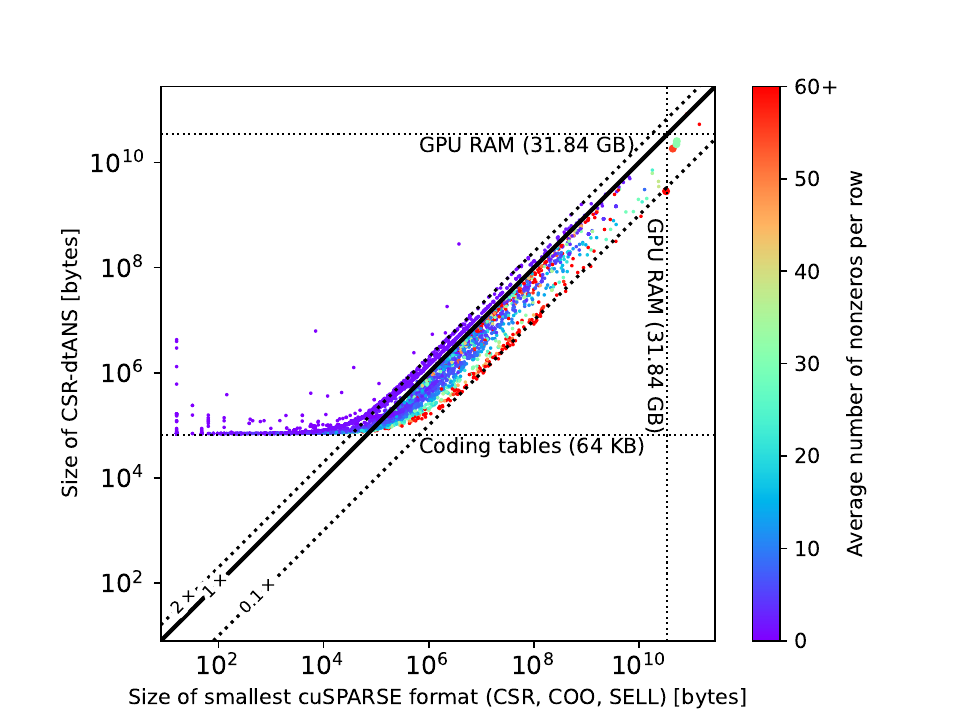}
\caption{64-bit}
\label{fig:results_compression_vs_cusparse_64}
\end{subfigure}
\begin{subfigure}{\linewidth}
\includegraphics[width=\columnwidth,trim={0 0 0 1.2cm},clip]{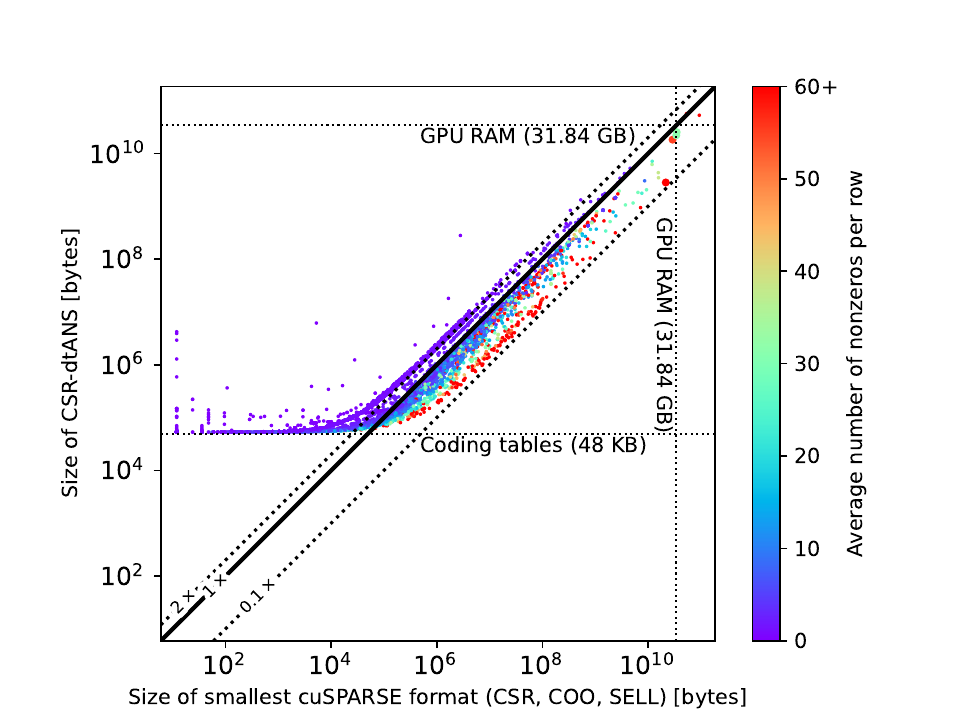}
\caption{32-bit}
\end{subfigure}
\caption{Matrix compression achieved with CSR-dtANS for 64- and 32-bit precision. Matrices below the solid line are compressed successfully.}
\label{fig:results_compression_vs_cusparse}
\end{figure}

We start by comparing the memory footprint of the considered matrices in CSR-dtANS to the smallest among CSR, COO, and SELL for 64- and 32-bit in Fig.~\ref{fig:results_compression_vs_cusparse}.

On the x-axis we plot the size in bytes for the most compressed version among COO, CSR and SELL, and on the y-axis we give the size in bytes for the same matrix compressed using CSR-dtANS. We plot the dots using different colors depending on the average number of nonzero elements per row, purple being a low number of nonzeros and red being a large number. In addition, we provide three diagonal lines. The solid line indicates equality between CSR-dtANS and the baseline. Points below the line correspond to successful compression by CSR-dtANS. The bottom, dotted line indicates an achieved compression ratio of 10$\times$; the top dotted line indicates a $2\times$ increase in size of CSR-dtANS compared to the baseline.

The outermost horizontal and vertical dotted lines indicate the amount of available GPU RAM; the horizontal line at 64 KB (64-bit)/48 KB (32-bit) is the constant size of the coding tables which equals $K=4096$ times the memory required per slot (16 bytes for 64- and 12 bytes for 32-bit).

From these measurements we observe two main results. First, dtANS is not suitable for small matrices, below $10^5$ bytes, say. The reason is the constant overhead of the coding tables. However, such small matrices are also not very important for GPU processing.

Second, dtANS better compresses matrices with sufficiently many nonzero elements per row. This is because a 4-byte word (to store the length \(n\) of the encoded sequence) is required regardless of the number of nonzeros of a row. This overhead is slowly amortized as more nonzeros are in a row. Notably, this is why a large group of matrices lies just above the 2x line: most of these matrices have only one nonzero per row and hence, around 4 words (1 for \(n\), 3 for \(w_1, w_2\), and \(w_3\)) are required instead of around 2 (for value and index in SELL format). Conversely, we see that all matrices approaching the $10 \times$ compression improvement have a high number of nonzero elements per row.
The best achieved compression is $11.77\times$ for 64-bit and $7.86\times$ for 32-bit. In both cases, the same 4 matrices (bold) fit in GPU RAM only with CSR-dtANS.

We study the relation between the number of nonzeros (total and per row) and successful compression further in Table~\ref{tab:results_compression_success_vs_cusparse}.

\begin{table}
\renewcommand{\arraystretch}{1.3}
\begin{subtable}{\linewidth}\centering
\begin{tabular}{@{}llll@{}}
\toprule& \multicolumn{3}{c}{\em total number of nonzeros in matrix}\\
\cmidrule{2-4}
\em annzpr & $\leq 2^{10}$ & $\leq 2^{15}$ & $>2^{15}$ \\
\midrule
$\leq 10$ & $0/1266=0$ & $1212/3868\approx0.31$ & $1119/1889\approx0.59$ \\
$>10$ & $0/14=0$ & $189/292\approx0.65$ & $1060/1061\approx1.00$ \\
\bottomrule
\end{tabular}
\caption{64-bit}
\label{tab:results_compression_success_vs_cusparse_64}
\end{subtable}
\ \\
\begin{subtable}{\linewidth}\centering
\begin{tabular}{@{}llll@{}}
\toprule& \multicolumn{3}{c}{\em total number of nonzeros in matrix}\\
\cmidrule{2-4}
\em annzpr & $\leq 2^{10}$ & $\leq 2^{15}$ & $>2^{15}$ \\
\midrule
$\leq 10$ & $0/1266=0$ & $908/3868\approx0.23$ & $1068/1889\approx0.57$ \\
$>10$ & $0/14=0$ & $168/292\approx0.58$ & $1060/1061\approx1.00$ \\
\bottomrule
\end{tabular}
\caption{32-bit}
\label{tab:results_compression_success_vs_cusparse_32}
\end{subtable}
\caption{Rate of successful compression with CSR-dtANS for 64- and 32-bit in relation to average number of nonzeros per row (annzpr) and total number of nonzeros in matrix.}
\label{tab:results_compression_success_vs_cusparse}
\end{table}

We group the matrices by their total number of nonzeros (columns) and by their average number of nonzeros per row (annzpr, rows). Each resulting cell states the fraction of successfully compressed matrices (but not the amount by which they were compressed). 

As observed before, compression is never successful for very small matrices with no more than $2^{10}$ nonzeros due to the constant overhead of the coding tables. For the others, a significant fraction could be compressed. The fractions increase with matrix size and annzpr and are higher for 64- than for 32-bit as expected. As a main result, with one exception, CSR-dtANS successfully compresses all 1061 matrices with at least $2^{15}$ nonzeros and at least 10 annzpr for both 64- and 32-bit.

\subsection{\spmvm runtime in warm-cache setting}

We benchmark our \spmvm on CSR-dtANS against the cuSPARSE kernel for the fastest format among COO, CSR, and SELL. First we consider a warm cache.

\begin{figure}
    \centering
\begin{subfigure}{\linewidth}
    \includegraphics[width=\linewidth,trim={0 0.8cm 0 3cm},clip]{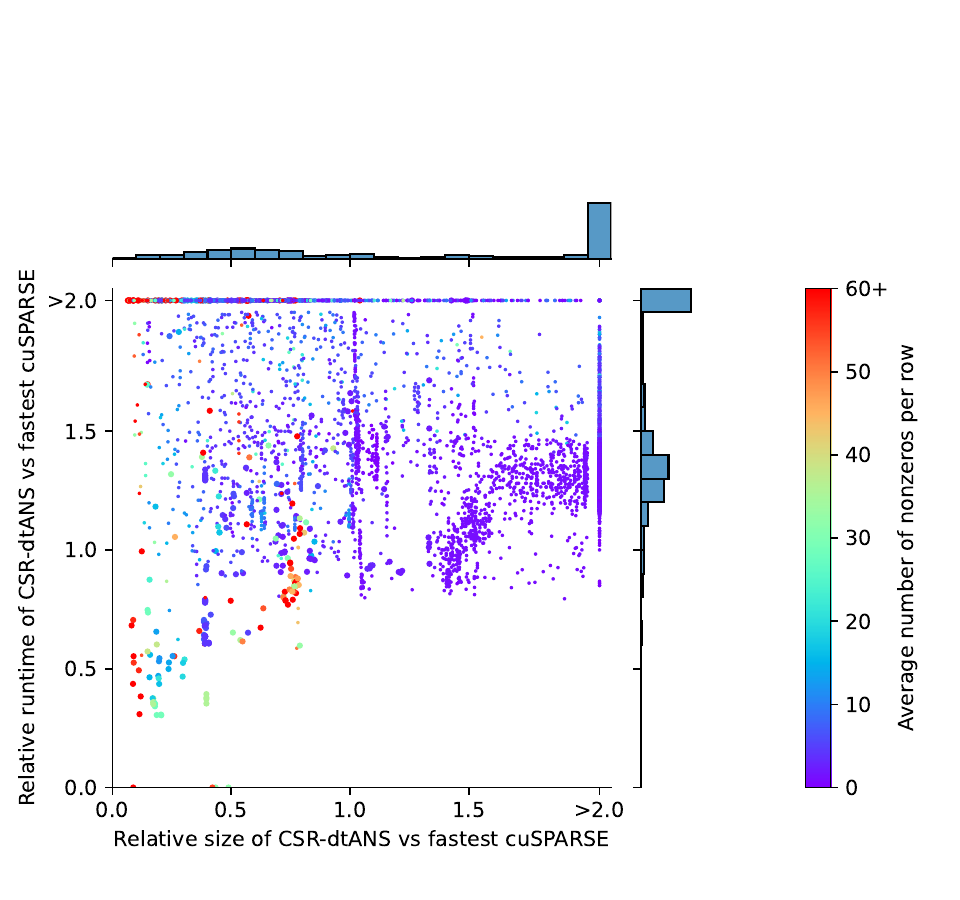}
    \caption{64-bit}
    \label{fig:results_speedup_vs_cusparse_warm_64}
\end{subfigure}
\begin{subfigure}{\linewidth}
    \includegraphics[width=\linewidth,trim={0 0.8cm 0 3cm},clip]{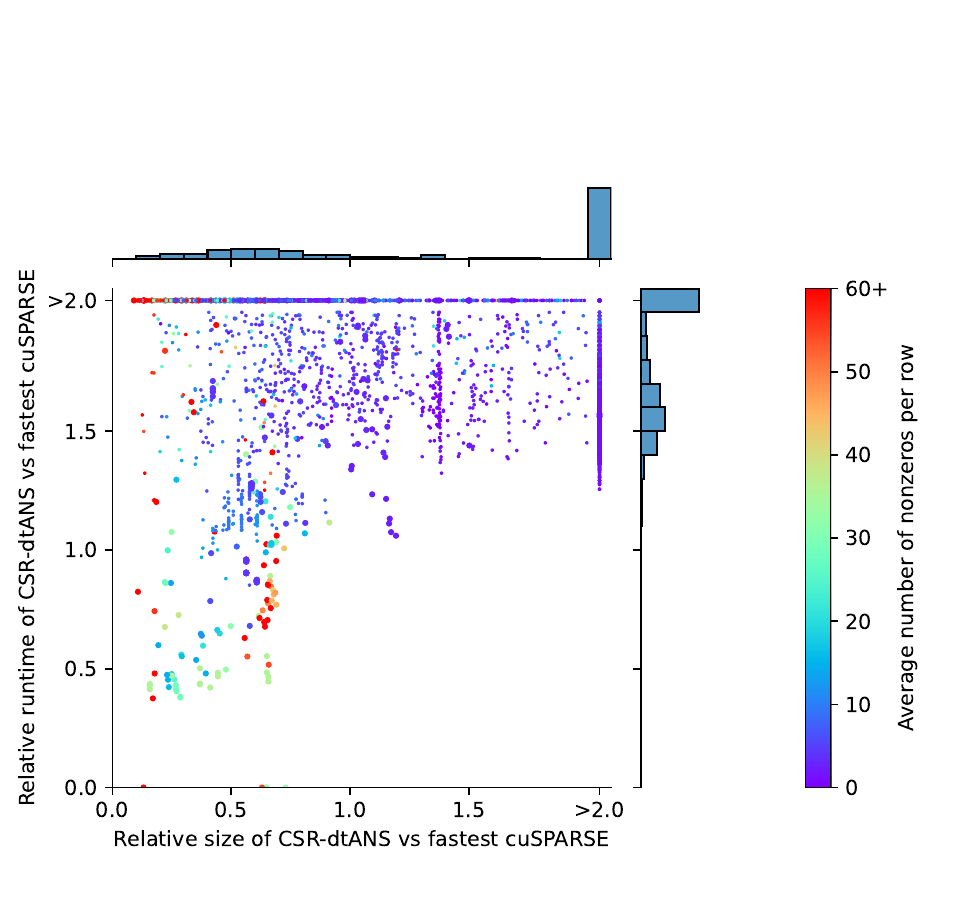}
    \caption{32-bit}\label{fig:results_speedup_vs_cusparse_warm_32}
\end{subfigure}
    \caption{Comparison of matrix size and \spmvm runtime against the fastest cuSPARSE format, run with warm cache, for 64- and 32-bit. Matrices in bottom half show a speedup. Bold matrices do \emph{not} fit in cache in fastest cuSPARSE format.}
    \label{fig:results_speedup_vs_cusparse_warm}
\end{figure}

Fig.~\ref{fig:results_speedup_vs_cusparse_warm} plots the relative runtime compared to the fastest baseline on the y-axis, so values below 1 (lower half) are a speedup for our \spmvm. On the x-axis we plot the relative size of CSR-dtANS compared to the fastest baseline. We use the same color scheme as before to indicate the average number of nonzeros per row of the matrix.

As expected, we generally observe that speedups with CSR-dtANS require the compression to succeed (bottom left quadrant). The few exceptions for 64-bit are likely due to an improved memory access pattern. Further, the highest speedups are achieved for matrices that were also compressed the most, and the speedup is less than the compression (i.e., practically all points lie above the diagonal in the bottom left quadrant). This is because plain CSR is also a baseline format. The maximum speedup is $3.29\times$ for 64-bit and $2.67\times$ for 32-bit (smaller, as expected by the lower compression factor). 

The converse is not true: many matrices are compressed efficiently but fail to show speedups (upper left quadrant). One cause is the fixed choice of CSR as a starting point for dtANS, which may be suboptimal. In particular, our \spmvm kernel does not handle well sparsity patterns with highly irregular number of nonzeros per row.
Lastly, there are also 4 dots on the x-axis itself, representing the 4 aforementioned matrices that only fit in GPU memory using CSR-dtANS.

As we did for successful compression before, we study the relation between the number of nonzeros (total and per row) and successful speedup with warm cache in Table~\ref{tab:results_speedup_success_vs_cusparse_warm}. This time columns focus on the largest matrices.

\begin{table}
\renewcommand{\arraystretch}{1.3}
\begin{subtable}{\linewidth}\centering
\begin{tabular}{@{}llll@{}}
\toprule& \multicolumn{3}{c}{\em total number of nonzeros in matrix}\\
\cmidrule{2-4}
\em annzpr & $\leq 2^{20}$ & $\leq 2^{25}$ & $>2^{25}$ \\
\midrule
$\leq 10$ & $217/6679\approx0.03$ & $23/264\approx0.09$ & $65/80\approx0.81$ \\
$>10$ & $0/913=0$ & $62/379\approx0.16$ & $40/75\approx0.53$ \\
\bottomrule
\end{tabular}
\caption{64-bit}
\label{tab:results_speedup_success_vs_cusparse_warm_64}
\end{subtable}
\ \\
\begin{subtable}{\linewidth}\centering
\begin{tabular}{@{}llll@{}}
\toprule& \multicolumn{3}{c}{\em total number of nonzeros in matrix}\\
\cmidrule{2-4}
\em annzpr & $\leq 2^{20}$ & $\leq 2^{25}$ & $>2^{25}$ \\
\midrule
$\leq 10$ & $0/6679=0$ & $2/264\approx0.01$ & $36/80\approx0.45$ \\
$>10$ & $0/913=0$ & $37/379\approx0.10$ & $42/75\approx0.56$ \\
\bottomrule
\end{tabular}
\caption{32-bit}
\label{tab:results_speedup_success_vs_cusparse_warm_32}
\end{subtable}
\caption{Rate of successful speedup with CSR-dtANS, run with warm cache, for 64- and 32-bit in relation to average number of nonzeros per row (annzpr) and total number of nonzeros in matrix.}
\label{tab:results_speedup_success_vs_cusparse_warm}
\end{table}

We achieve almost no speedups for small matrices with up to $2^{20}$ nonzeros. Up to $2^{25}$ nonzeros (next column) first speedups appear, in more than 10\% of the cases for the higher annzpr. For the largest, and arguably most relevant, matrices, we see speedups for the majority of matrices. 

Generally, speedups are more likely for 64-bit as expected, but for the annzpr the connection is less clear. Note that the regularity and structure of the matrix also play an important role but are difficult to express quantitatively.

We conclude that size is the most important feature to predict whether a matrix is likely to see a speedup; whereas the number of nonzeros per row determines the magnitude of it (as seen in Fig.~\ref{fig:results_speedup_vs_cusparse_warm}). 

\subsection{\spmvm runtime in cold-cache setting}

We continue by repeating the prior benchmark, this time run with cold cache, which should favor compression better than warm cache.

\begin{figure}
    \centering
\begin{subfigure}{\linewidth}
    \includegraphics[width=\linewidth,trim={0 0.8cm 0 3cm},clip]{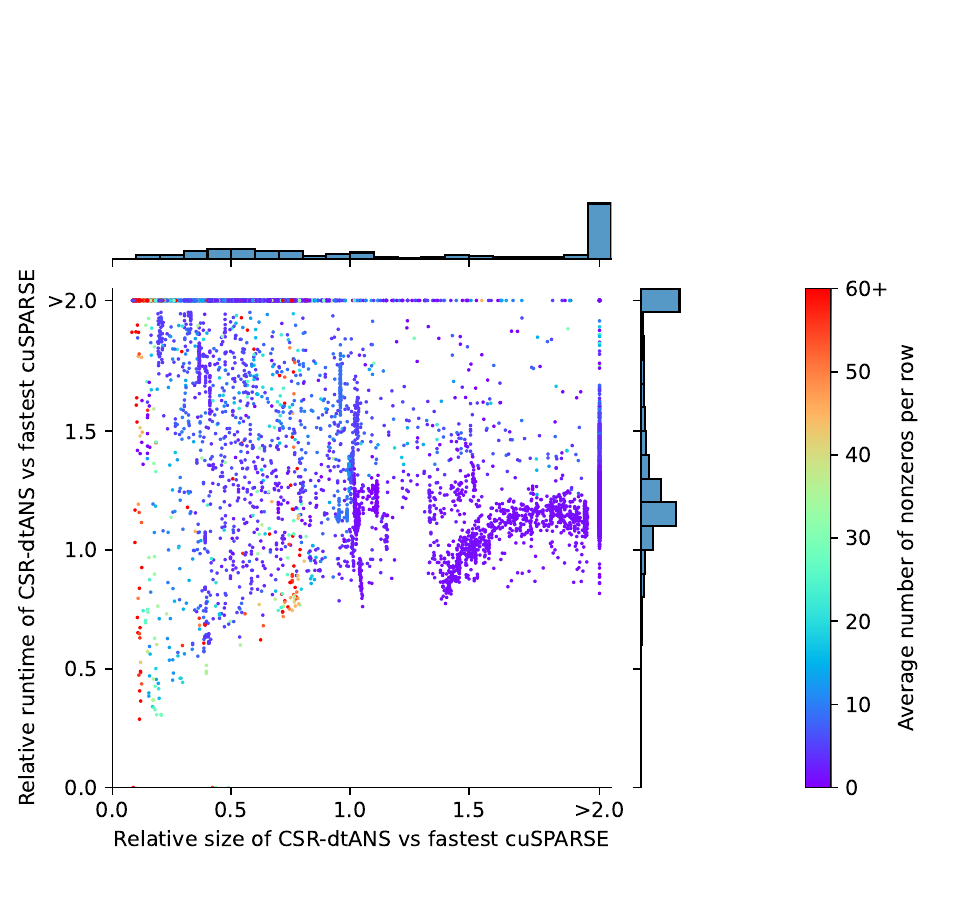}
    \caption{64-bit}
    \label{fig:results_speedup_vs_cusparse_cold_64}
\end{subfigure}
\begin{subfigure}{\linewidth}
    \includegraphics[width=\linewidth,trim={0 0.8cm 0 3cm},clip]{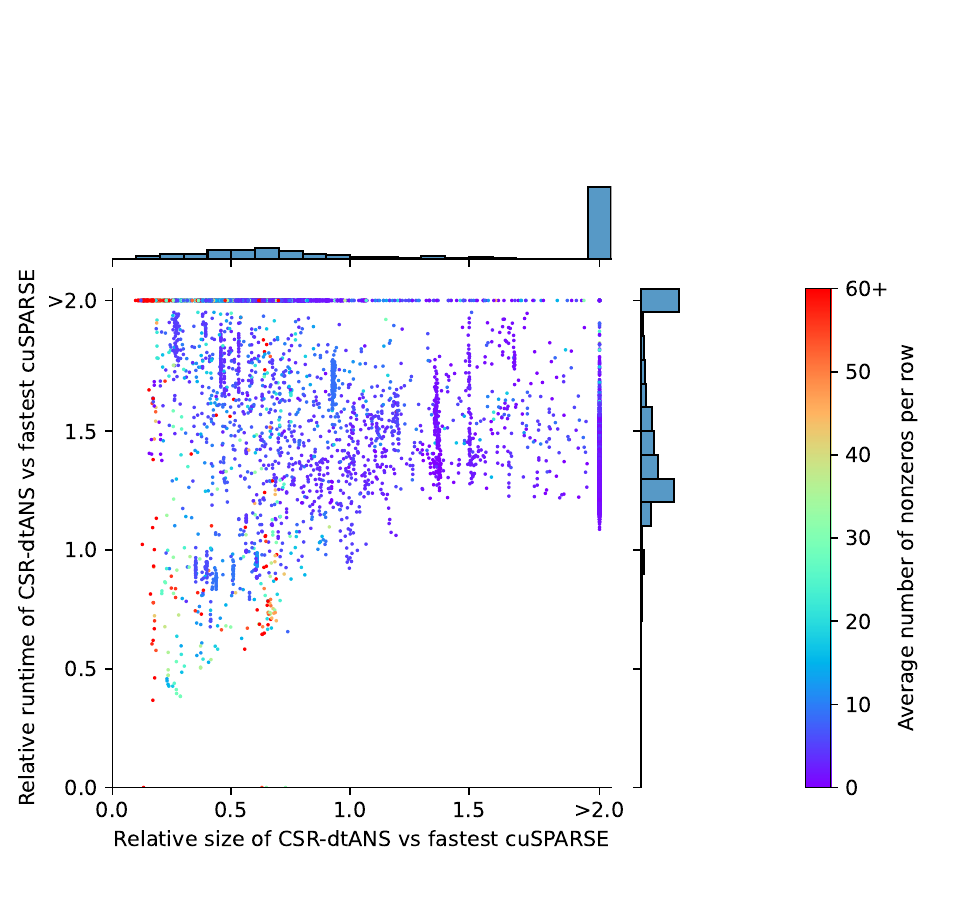}
    \caption{32-bit}\label{fig:results_speedup_vs_cusparse_cold_32}
\end{subfigure}
    \caption{Comparison of matrix size and \spmvm runtime against the fastest cuSPARSE format, run with cold cache, for 64- and 32-bit. Matrices in bottom half show a speedup.}
    \label{fig:results_speedup_vs_cusparse_cold}
\end{figure}
Fig.~\ref{fig:results_speedup_vs_cusparse_cold} plots relative runtime against relative matrix size compared to baseline analogously to the warm-cache results in Fig.~\ref{fig:results_speedup_vs_cusparse_warm}.

The main observations are qualitatively identical to the warm-cache setting. One minor change is that the points now almost all lie above the diagonal of the bottom left quadrant. Further, we observe higher speedups for more matrices as we quantify further below. The maximum speedup is $3.48\times$ for 64-bit and $2.72\times$ for 32-bit. 

As we did for warm cache in Table~\ref{tab:results_speedup_success_vs_cusparse_warm}, we study the relation between the number of nonzeros (total and per row) and successful speedup with cold cache in Table~\ref{tab:results_speedup_success_vs_cusparse_cold}.

\begin{table}
\renewcommand{\arraystretch}{1.3}
\begin{subtable}{\linewidth}\centering
\begin{tabular}{@{}llll@{}}
\toprule& \multicolumn{3}{c}{\em total number of nonzeros in matrix}\\
\cmidrule{2-4}
\em annzpr & $\leq 2^{20}$ & $\leq 2^{25}$ & $>2^{25}$ \\
\midrule
$\leq 10$ & $374/6679\approx0.06$ & $62/264\approx0.23$ & $64/80\approx0.80$ \\
$>10$ & $3/913\approx0.00$ & $109/379\approx0.29$ & $46/75\approx0.61$ \\
\bottomrule
\end{tabular}
\caption{64-bit}
\label{tab:results_speedup_success_vs_cusparse_cold_64}
\end{subtable}
\ \\
\begin{subtable}{\linewidth}\centering
\begin{tabular}{@{}llll@{}}
\toprule& \multicolumn{3}{c}{\em total number of nonzeros in matrix}\\
\cmidrule{2-4}
\em annzpr & $\leq 2^{20}$ & $\leq 2^{25}$ & $>2^{25}$ \\
\midrule
$\leq 10$ & $117/6679\approx0.02$ & $35/264\approx0.13$ & $37/80\approx0.46$ \\
$>10$ & $4/913\approx0.00$ & $132/379\approx0.35$ & $46/75\approx0.61$ \\
\bottomrule
\end{tabular}
\caption{32-bit}
\label{tab:results_speedup_success_vs_cusparse_cold_32}
\end{subtable}
\caption{Rate of successful speedup with CSR-dtANS, run with cold cache, for 64- and 32-bit in relation to average number of nonzeros per row (annzpr) and total number of nonzeros in matrix.}
\label{tab:results_speedup_success_vs_cusparse_cold}
\end{table}

As expected, the success numbers are uniformly increased compared to Table~\ref{tab:results_speedup_success_vs_cusparse_cold}, since now all loads are from RAM without the help of caches. The greatest improvement is for matrices with less than $2^{25}$ (and more than $2^{20}$) nonzeros. For the largest matrices, the change is small compared to warm cache. The main reason is that for those the cache state makes less of a difference since they mostly do not fit into cache anymore. E.g., even a matrix with $2^{25}$ nonzeros requiring more than 12 bits per nonzero would already exceed the 96~MB cache. Another effect may be the irregularity of the row lengths, which is punished more in a warm-cache setting.

In summary, the cold-cache setting reduces the matrix size threshold that determines whether CSR-dtANS might be a viable option.

\subsection{Speedup vs.~AlphaSparse}

\begin{figure}
\centering
\includegraphics[width=\columnwidth,trim={0 0.8cm 0 3.5cm},clip]{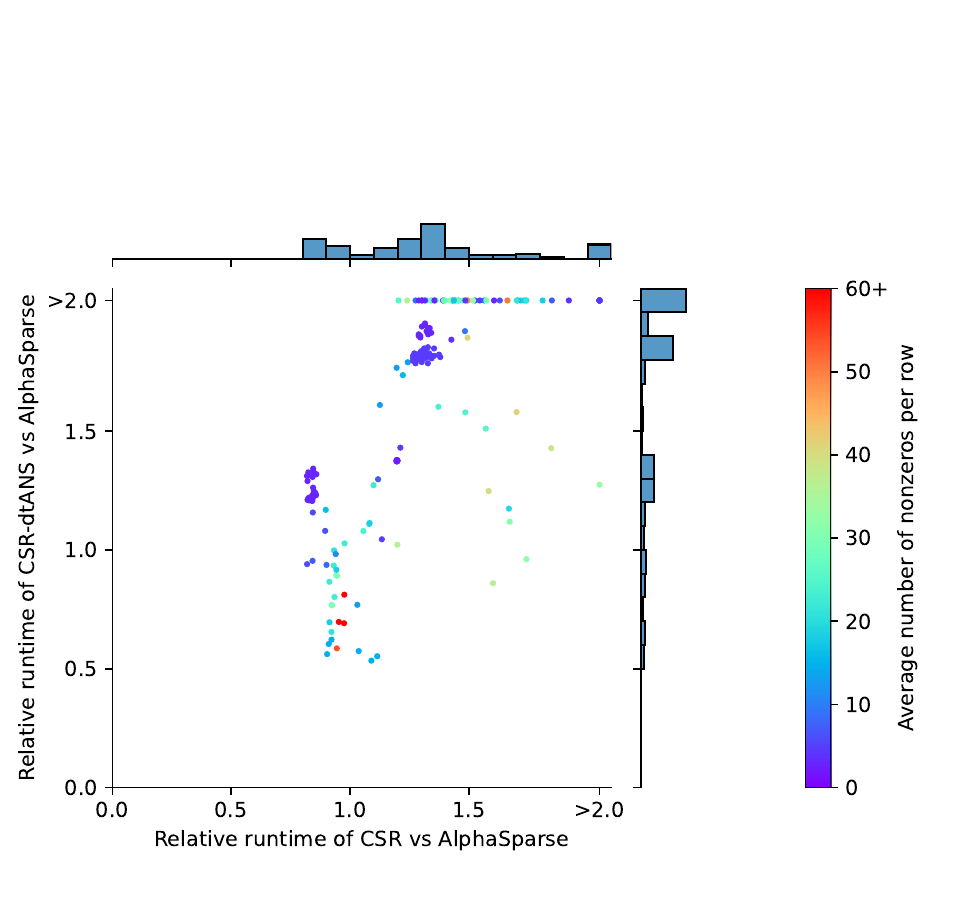}
\caption{Comparison of CSR and CSR-dtANS \spmvm runtime against AlphaSparse (warm cache, 32-bit, no duplication of off-diagonal elements for symmetric matrices).}
\label{fig:results_speedup_vs_alphasparse}
\end{figure}

Finally, we compare the runtime of the evaluated matrices in CSR-dtANS to AlphaSparse in Fig.~\ref{fig:results_speedup_vs_alphasparse}. We now use the warm-cache setting, 32-bit precision and handle symmetry as AlphaSparse does (explained earlier in  this section). Due to the extreme cost of executing AlphaSparse (several hours per matrix for autotuning), we evaluate it only for a promising family of 265 matrices. This family was selected as those matrices for which CSR-dtANS improved in size and runtime by at least 10~\% over the best cuSPARSE format in an early experiment. For 36 of these, AlphaSparse failed to produce an output kernel. Fig.~\ref{fig:results_speedup_vs_alphasparse} shows the results for the remaining 229. The plot is almost analogous to the ones before. However, on the x-axis we plot the relative runtime of CSR to AlphaSparse instead of the relative size. Technically, this should result in all matrices lying in the right half, because CSR is contained in the space explored by AlphaSparse. In spite of that, 52 matrices lie in the left half, hinting that AlphaSparse failed to find the optimal kernel for those. Furthermore, due to the choice of matrices, all data points should be below the diagonal. However, this is not the case because CSR-dtANS was faster at the time the matrices were selected than the current version (it did not yet support escaping values).

Despite CSR-dtANS being a fixed format and AlphaSparse autotuning for several hours, we observe a speedup for 28 matrices, some of which are quite large, and up to $1.87\times$.

\section{Related Work}
\label{sec:related_work}
In memory-bound workloads, performance is limited primarily by data transfer between main memory and the processor.
Reducing memory traffic through compression, even at the cost of additional computation, can yield significant speedups. A large corpus of work aims to improve the performance of \spmvm on both CPU and GPU. In this section, we provide a selection of related work on compression for \spmvm.

\mypar{\spmvm on CPU}
Early work on CPUs adopted the natural strategy of omitting zero entries, as implemented in the classical compressed sparse row (CSR) and coordinate (COO) formats~\cite{saab1994sparskit, Im2004}, which remain the basis for modern linear algebra libraries.
Some works focus on designing specialized formats optimized for data locality and SIMD utilization~\cite{wang2024, spmv2002}, while others use code generation techniques for specific sparsity patterns~\cite{cheshmi2022}. Other approaches aim to improve compression ratios by exploiting repeated patterns~\cite{patternspmv2009} or recurring indices~\cite{Willcock2006,Kourtis2010}. Some projects push this compression further by applying delta or run-length encoding to indices to reduce bandwidth consumption~\cite{galanopoulos2025,donenfeld2022}.

\mypar{\spmvm on GPU}
Classical CPU-oriented sparse matrix formats were not designed to exploit thread- and warp-level parallelism. Early GPU implementations of standard formats revealed significant inefficiencies due to thread divergence and uncoalesced memory accesses~\cite{bell2009spmvgpu}. These limitations motivated the development of specialized GPU formats that improve workload balancing and memory access efficiency across warps~\cite{vazquez2010spmvgpu, Monakov2010, Ashari2014, greathouse2014spmvgpu, Liu2015, Anzt2014spmvgppu}. Other approaches introduced auto-tuning frameworks~\cite{choi2010spmvgpu, Su2012}, machine learning-based format selection methods~\cite{sedaghati2015spmvgpu, Du2022, Zhao2018}, and algorithmic optimization methods~\cite{niu2021spmvgpu}, to choose the best format or combination of formats to represent a sparse matrix. Modern approaches leverage hardware-specific instructions to improve performance~\cite{ma2025spmvgpu}.

\mypar{\spmvm using entropy coding}
To our best knowledge, we are the first to apply general entropy coding for \spmvm. However, some works explore variable bit-length delta-encoding~\cite{lawlor2013, Trommer2021}. In contrast to true entropy coding, these approaches use a fixed mapping from deltas to codewords assigning small deltas to short codewords, an assumption that might not provide good compression for all sparsity patterns.

\section{Conclusion}

Our main contribution is to show that a generic entropy encoding technique, our proposed dtANS, can provide fast enough on-the-fly decoding to speed up a memory-bound computation on GPUs. We demonstrated such speedups for \spmvm against the strong cuSPARSE baseline, even including warm-cache scenarios. There are various avenues to further performance gains including assigning several rows to each thread (which would remove our limitation for short rows) or using formats other than CSR as a starting point for entropy coding. We believe that hardware trends~\cite{memwall2024} will further increase the applicability of our approach for 
\spmvm and will also make it a candidate for other memory-bound computations on GPUs.


\section{References}
\printbibliography[heading=none]

\end{document}